\documentclass{llncs}
\pdfoutput=1
\usepackage{graphicx}
\usepackage{amsfonts, amsmath,amscd}
\usepackage{mathtools}

\title{A General Theory of Information and Computation}
\author{Pieter Adriaans\inst{1}}

\institute{ILLC, FNWI-IVI, SNE\\
University of Amsterdam,\\Science Park 107\\ 1098 XG Amsterdam, \\The
Netherlands.
\email{P.W.Adriaans@uva.nl}}

\date{}
\begin{document}
\maketitle

\begin{abstract}
This paper fills a gap in our understanding of the interaction between information and computation. It unifies other approaches to measuring information like Kolmogorov complexity and Shannon information. We define a theory about information flow in deterministic computing based on three fundamental observations: 1) $\forall (n \in \mathbb{N}) I(n)=\log n$ (information is measured in logarithms),  2) all countable sets  contain the same amount of information and  3) $\forall (x) I(x) \geq I(f(x))$ (deterministic computing does not create information). We analyze the flow of information through computational processes: exactly, for primitive recursive functions and elementary artithmetical operations and, under maximal entropy, for polynomial functions and diophantine equations. Thus we get, by the MRDP-theorem, a theory of flow of information for general computable functions.  We prove some  results like the Fueter-P\'{o}lya conjecture and the existence of an information conserving enumeration of all finite sets of numbers. We also show that the information flow in more complex derivatives of the primitive recursive functions like addition and multiplication is not trivial: in particular associativity is not information efficient for addition. Using the Cantor pairing function we develop a universal measuring device for partitions of the set of finite sets of numbers. We show that these sets can be enumerated by a polynomial function when ordered by cardinality, but not when ordered by their sums. 
\end{abstract}

\newpage
\section{ Summary of Main Proof}
We give a short simplified overview of the proof of one of the main consequences of the theory presented in the paper: 

\begin{theorem}The set of sets of natural numbers is countable when ordered by cardinality but not countable when ordered by sums. 
\end{theorem}
We compute to transform existing information in to new information and some computations are easier than others. We know this from everyday experience.~\footnote{I thank Rini Adriaans for this example.} Suppose we want to add the numbers $2$, $47$, $53$ and $98$. Most of us will see the pattern that makes this easy: $(2+98) + (47 + 53) = 100 + 100 = 200$. The numbers have a special relationship that makes the result is less surprising, and therefore less informative. An interesting consequence of the general theory is that, even for addition, there are an infinite number of these tricks and there is no way to investigate them systematically. 

The structure of the proof is as follows.  We consider functions on the natural numbers. If we measure the amount of information in a number $n$ as: \[I(n) = \log n\] then we can measure the information effect of applying function $f$ to $n$ as: \[I(f(n)) = \log f(n)\] This allows us to estimate the \emph{information efficiency} as: \[\delta(f(n)) = I(f(n)) - I(n)\] Consequently, for large enough numbers, there are functions that are information discarding, like addition ($\log (a + b) - \log a - \log b < 0$), information conserving like multiplication ($\log (ab) - \log a -\log b = 0$) and information expanding like exponentiation ($\log x^5 - \log x >0$).  

We can now explain the mathematical aspect of the phenomenon described above. The conditional information in the number $200$  given the set $\{2, 47, 53, 98\}$ varies with the method of computation:   $\delta(2+98) + \delta(47 + 53) + \delta(100  + 100)  \approx -1.08 > \delta(2 +  47) + \delta(53 +  98) + \delta(49 + 151)  \approx -1.74$. In short: \emph{information efficiency is non-associative for addition}. 

The fact that a set adds up to a certain number, gives us information about the set, but the amount of information we gain or lose when doing the computation varies. We analyze this phenomenon for the class of all possible additions of sets of numbers. A useful tool is the:

 \begin{definition}[Law of Conservation of Information]\label{INFCONS}

 If $S$ is a countable infinite set $f$ and $g$ are total functions on $S$ such that there is a third function  $\pi(f(x),g(x))$ that defines a bijection between $S$ and the natural numbers then in the limit for each $s \in S$: 
\[I(s) = I( \pi(f(x),g(x))) = I(f(s)) +I (g(s)) = -\delta(f(x) - \delta(g(x))\]
\end{definition}
This means that $\pi$ is an information conserving function, while $f$ and $g$ are information discarding, but in perfect balance with the information generated by $\pi$. The law of conservation of information allows us to split the set $S$ in to an infinite amountof countable partitions, for which we know the exact information: $f$ gives the partitions and $g$ gives the index of elements in the partitions. The function $\pi(f(x),g(x)))$ defines an information grid that functions as a measuring tool. 

We show that the set of sets of finite natural numbers is measurable in this way: $f$ gives the cardinality of the sets, $g$ enumerates the sets of cardinality $k$ lexicografically as a combinatorial number system, $\pi$ is the Cantor pairing function. We say that  \emph{the set of finite sets of numbers is countable on cardinalities}: each set has a location in a two dimensional discrete grid where the columns contain the sets with the same cardinality and the rows give the indexes of these sets. These bijections are computable in polynomial time. We actually prove that this is the only possible enumeration of the set of finite sets of numbers, but this result is not essential for the following observation.

Suppose $\Sigma$ is the function that sums up the elements of a set of numbers, $\theta$ is the function that indexes elements of sets with the same sums and  $\eta(\Sigma s,\theta_{\Sigma s}(s))$ defines a bijection to the natural numbers. The functions must obey the law of conservation of information: 

\[I(s) = I( \eta(\Sigma s, \theta(s))) = I(\Sigma s) +I (\theta(s)) = - \delta(\Sigma s) - \delta(\theta(s))\]

Note that, since we already assume that the set is algorithmically countable, the precise definition of $\eta $ is not relevant here: $\Sigma s$ and $\theta(s)$ must contain the exact amount of information to reconstruct $s$ and vice versa, moreover $\theta$ is one function with a defined information efficiency, for which for each $s$ we have exactly:  $I(s) = - \delta(\Sigma s)) - \delta(\theta(s))$. This is not possible because $\Sigma$ is not one function, but an infinite set of functions, for which the information efficiency varies unboundedly, both over cardinalities of sets as within partitions of sets with the same cardinality: \emph{the set of finite sets of numbers is not countable  on sums}.  By similar reasoning one can prove that  \emph{the set of finite sets of numbers is not countable  on products.}

Summarizing: 
\begin{itemize}
\item There is a computable procedure that gives us the $n$-th element of a set of sets of numbers with cardinality $k$.
\item There is no computable procedure that gives us the $n$-th element of a set of sets of numbers that add up to $k$ or that multiply up to $k$.  
\end{itemize}
This has consequences for search procedures. Concepts like \emph{the first subset of a set that adds up to $k$ or multiplies to $k$} have no meaning. The only context in which the notion of \emph{the first set} makes sense is when the sets are ordered on cardinality. Any systematic search needs to use the standard enumeration via the cardinalities of the sets.  Sums and products of sets give us no information that helps us to identify them. This first observation provides, via the subset sum problem, a strategy to prove a separation between the classes $P$ and $NP$ the second suggests that there is no general method for factorization. 

In the following paragraphs we describe a general theory of information and computation that supports these results. 

\newpage

\section{A General Theory of Information and Computation}
\subsection{Introduction}

The theoretical constructions by which we conclude that one object contains information about an other object are  diverse: if we consider cattle in Mesopotamia around 1500 B.C. the collection of objects could be a set of clay balls of various shapes in an urn, and the narrative could be that the balls correspond to sheep or goats of various sex and age in a heard. We can measure the temperature in a room in terms of the length of a column of mercury given that the column is calibrated and constructed in a certain way. We can identify an individual by means of records in databases, which are ultimately configurations of bits on a memory chip and via a complex process come to us as understandable data. 

These stories specify the semantic aspect of information in a certain situation. Given the richness of everyday life it is unlikely that a comprehensive theory of various types of narratives can ever be formulated, but in mathematical sense such a construction is just a function: 

\begin{definition}
A collection of objects $D$ (abstract or concrete, continuous or discrete, finite or infinite) contains information about an element $s$ of a set of objects $R$ (abstract or concrete, continuous or discrete, finite or infinite) in a certain context if there is a function that allows us to identify a proper subset of $S$ that contains $s$. 
\end{definition}

Abstracting from narratives (gases in cylinders, systems of messages, sets of strings, columns of mercury, length of rods, sets of Turing machines, true well-formed statements, probabilities, records in databases etc.) we can develop a theory of information in the foundational language of mathematics: set theory and arithmetic. A basic insight is: 

\begin{definition}[Information in elements of Sets]\label{INfFORMATIONINSETS}
The information that predicate $P$ applies to object $s$ is equivalent to the information that $s$ belongs to the set of things that are $P$.
\end{definition}

Our intuitions in every day life about the relation between information and sets are unclear and sometimes contradictory. Information measures are non-monotone over set theoretical operations. In the majority of cases for any information measure $I$ of sets, if $B \subset A$, then $I(A) > I(B)$ if $A$ is  sparse and $I(A) < I(B)$ if $B$ is sparse. Clarification of these issues is one of the main objectives of this paper.  

\subsection{Counting and Measuring in the Limit}
In nature we distinguish deterministic processes, for which at any moment the next step has only one fixed choice, and non-deterministic processes for which the future development is open. The essence of measurement is predictability and repeatability.  Counting is a deterministic process that establishes a one-to-one correspondence between spatial representations of numbers and the \emph{cardinality} of a collection of objects. Measuring is a process establising a one-to-one correspondence between representations of numbers and the \emph{length} of an object. 
 
 The most elementary counting process is tallying, associated with a unary number system. Here the number representations are homogeneous, with low entropy, and the length of the representation is equivalent to the cardinality of the set that is counted. If we count using a positional system then the number representations are heterogeneous, in general with high entropy.  The length of the representation is equivalent to the logarithm of the unary number of the cardinality of the set that is counted. 
 Given a number $r$ with a positional representation of length $n$ we have  $n = \lceil \log (r) \rceil$. It seems natural to take this value, that is associated with the length of a representation, as a measure of the amount of information in a number, so, abstracting from discrete representations, we define:~\footnote{The base of the logarithm is of no specific importance. If one chooses the natural logarithm, information is measured in \emph{gnats} if we choose base two the measure is in \emph{bits}, but any other base is acceptable. We will use the expression $\log_a$ of $\log$ for any general logarithm with base $a$ and $\ln$ for the natural logarithm.}
 
 \begin{definition}[First Information Law of Deterministic Computing]\label{FIRSTLAW}
\[\forall (x \in \mathbb{N}) I(x)=\log x\]
\end{definition}
Note that $n \in \mathbb{B}$ is a \emph{cardinality} and that the measurements $I(n)$ represent a cloud of \emph{scalar} point values in $\mathbb{R}$. The relation between the two sets is described by the Taylor series for $\log (x+1)$:   
\begin{equation}\label{MEASINFGROW}
I(s+1) = \log (x + 1) =  \Sigma_{n=1}^{\infty}(-1)^{n-1}\frac{x^n}{n} \end{equation}

We have $\lim_{x \rightarrow \infty} I(x+1) - I(x) = 0$. In the limit the production of information by counting processes stops as specified by the derivative $\frac{d}{dx} \ln x = \frac{1}{x}$. The intuition that one could express based on this observation is that information in numbers in the limit becomes extremely dense, much denser than the values in `normal' counting processes. This implies that in the limit we have an unlimited amount of random numbers with the same information at our disposal.~\footnote{See Appendix \ref{SETSINCOMP} for a discussion.} Limits on the information in numbers are more well behaved than standard limits.~\footnote{A more extensive analysis of information measurement in the limit is given in appendix \ref{TRANSFININFMAN}.} In general we may conclude that since the log operation reduces multiplication to addition of exponents of variables, in the limit it looses its sensitivity to addition in the variables itself. In many cases limit statements about the amount of information in variables are easier to prove than limit statements about the values of the variables. This gives us a tool to analyse the interplay between computation and information. It is well known that the set of natural numbers is incompressible: in the limit all sets that are compressible by more than a constant factor have density $0$.~\footnote{See section \ref{INFINSSETS} for a discussion.} This provides us with the: 

 \begin{definition}[Second Information Law of Deterministic Computing]\label{SECONDLAW}
All countable sets contain the same amount of information. 
\end{definition} 
All we have to do to measure information in an infinite set is define a bijection with the natural numbers. Given two sets $D$ and $R$ we say that a total function $f: D \rightarrow R$ 
\begin{itemize}
\item \emph{carries no information} about an element $s  \in D$ if $f^{-1}f(s)=D$, 
\item it \emph{identifies} $s$  if $f^{-1}f(s)=\{s\}$, and it 
\item \emph{carries} information if $f^{-1}f(s) \subset D$. 
\end{itemize}
Consequently functions are interesting from an information point of view if they map the natural numbers to proper subsets: i.e. they expand information. Many mathematical functions actually expand information (e.g. $f(x) = x^5$). They generate more output then their input `justifies'. By definition this newly generated information must be vacuous, i.e. the output generated by the function is compressible. In this paper we will consider functions that are defined by deterministic processes. From a philosophical point of view we cannot get new information about things that we do already know. Consequently, if we know  the future of a deterministic process, then this process cannot generate information. This leads to the: 

 \begin{definition}[Third Information Law of Deterministic Computing]\label{THIRDLAW}
\[\forall (x) I(x) \geq I(f(x))\]
\end{definition} 
Deterministic computation can conserve information, it can discard information but it cannot create information. This statement is not an axiom but it can actually be proved. A proof for Turing machines is presented in \cite{AB2011}. A related insight from information theory is the so-called data processing inequality that states that information is in general lost and never gained when transmitted through a noisy channel.~\cite{CT2006}

\subsection{Information in Infinite Sets of Natural Numbers}\label{INFINSSETS}

A set $S$ is Dedekind infinite if there is a bijective function onto some proper subset $A$ of $S$. In this sense the set of natural numbers $\mathbb{N}$ is infinite. 
 A subset $A$ of $\mathbb{N}$ is by definition countable via its index function.

\begin{definition}[Conditional Information]\label{CONDITIONALINFORMATION}
Let A be a subset of the set of natural numbers $\mathbb{N}$. For any $n \in \mathbb{N}$ put $A(n)=\{1,2,\ldots,n\} \cap A$:
\begin{itemize} 
\item The \emph{index function} of $A$ is $i_A(j)=n$, where $n=a_j$ the $j$-th element of $A$.  
\item the \emph{conditional information}  in $n$ given $A$ is  $I(n|A) = \log j$, where  $i_A(j)=n$.
\item the \emph{randomness deficiency}~\footnote{For historical reasons we will call this notion \emph{randomness deficiency}. It is related to a similar notion in Kolmogorov complexity \cite{LiVi08}}  generated by $A$ for $n$ is  $\delta_A(n) = \log n - \log j$, where  $i_A(j)=n$. 
\item The \emph{compression function} of $A$ is $c_{A}(n)=|A(n)|$.
\end{itemize}
\end{definition}

The density of a set is defined if in the limit the distance between the index function and the compression function does not fluctuate too much: 

\begin{definition}
Let A be a subset of the set of natural numbers $\mathbb{N}$ with  $c_{A}(n)$ as compression function. The lower asymptotic density $\underline{d}(A)$ of $A(n)$ in $n$ is defined as: 

\begin{equation}
\underline{d}(A) = \liminf_{n \rightarrow \infty}  \frac{c_{A}(n)}{n} 
\end{equation}

The upper asymptotic density  $\overline{d}(A)$ of $A(n)$ in $n$ is defined as: 

\begin{equation}
\overline{d}(A) = \limsup_{n \rightarrow \infty}  \frac{c_{A}(n)}{n} 
\end{equation}

The natural density  $d(A)$ of $A(n)$ in $n$ is defined when both the upper and the lower density exist as: 

\begin{equation}
d(A) = \lim_{n \rightarrow \infty} \frac{c_{A}(n)}{n} 
\end{equation}
\end{definition}

The density of the even numbers is $\frac{1}{2}$. The density of the primes is $0$. Note that $d$ is not a measure. There exist sets $A$ and $B$ such that $d(A)$ and $d(B)$ are defined, but $d(A \cup B)$ and $d(A \cap B)$ are not defined. The set of decimal numbers that start with a $1$ has no natural density in the limit. It fluctuates between $1/9$ and $5/9$ indefinitely. The sequence $ A=\{a_1<a_2<\ldots<a_n<\ldots; n\in\mathbb{N}\}$  has a natural density $\alpha \in [0,1]$ iff  $d(A) = \lim_{n \rightarrow \infty} \frac{n}{c_{A}(n)}$ exists.~\cite{NIVEN51} The notions of density of a subset of $\mathbb{N}$, compressibility and information are closely associated. Suppose $A$ is the set of primes. By the prime number theorem the index of a prime in $A$ will be in the limit close to $e^{\ln(n) - \ln(\ln(n))}$, which contains $\ln e^{\ln(n) - \ln(\ln(n))} = \ln \frac{x}{\ln x}$ information. Knowing that a number is prime reduces the amount of information we need to identify the number by a logarithmic factor.  

A useful notion is:

\begin{definition}[Information Efficiency Function]\label{RANDDEFFUNC}
When the density of a set $A$ is defined then the \emph{randomness deficiency function} $f$ is defined by: 
\[\lim_{n\to\infty}\frac{c_{A}(n) f(n)}{n}=1\]   
\end{definition}
 A natural interpretation for this expression is that $\log c_{A}(n)= I(n|A)$ is a measure of the amount of information in the nunber $n$, given the set $A$, $\log n= I(n)$ is the amount of information in the number $n$ without knowing $A$, and $\log f(n) = I(n) - I(n|A)$ is the information we gain by knowing $A$. The following theorem shows that the amount of compressible numbers is very limited: 

\begin{theorem}[Incompressibility of the set $\mathbb{N}$] \label{INCOMPRESSIBILITY}
The density of the set of numbers compressible by more than a constant factor is zero in the limit.
\end{theorem} 
Proof: Suppose $A$ is the set of numbers compressible by more than any constant in the limit. Assuming the density of $A$ is defined we have:  $\log c_{A}(n) + \log f(n) = \log n$.  Here $\log f(n)$, where $f$ is the randomness deficiency function, is a measure for the information we gain knowing that $n \in A$. By definition $\log f(n)$ is not bounded by a constant in the limit. So we have:
 \[d(A) = \lim_{n\to\infty}\frac{c_{A}(n)}{n}=\frac{1}{f(n)}=0\] 
Suppose the density is not defined. By the result described above for those points where $\overline{d}(A)$ is realized it also goes to zero in the limit, since the compression of elements of $S$ is not bounded by a constant. But then since $0 \leq \underline{d}(A) \leq \overline{d}(A) \leq 0$. We have, contrary to the assumption, that the density of $A$ is defined and $d(A)=0$.  $\Box$

\subsection{Information Efficiency of Functions}
We can now specify more exactly what it means for a function to `carry information'. The \emph{Information Efficiency} of a function is the difference between the amount of information in the input of a function and the amount of information in the output. We use the shorthand $f(\overline{x})$ for  $f(x_1,x_2,\dots,x_k)$: 

\begin{definition}[Information Efficiency of a Function]\label{EFFFUNCTION}
Let $f: \mathbb{N}^k \rightarrow \mathbb{N}$  be a function of $k$ variables.  We have:
\begin{itemize} 
\item the \emph{input information} $I(\overline{x})$ and 
\item the \emph{output information}   $I(f(\overline{x}))$. 
\item The information efficiency of the expression $ f(\overline{x})$ is  
\[\delta(f(\overline{x}))= I(f(\overline{x})) - I(\overline{x})\]
\item A  function $f$ is \emph{information conserving} if $\delta(f(\overline{x}))=0$ i.e. it contains exactly the amount of information in its input parameters, 
\item it is \emph{information discarding} if  $\delta(f(\overline{x}))<0$ and 
\item it has \emph{constant information } if  $\delta(f(\overline{x})) = c$. 
\item it is \emph{information expanding} if  $\delta(f(\overline{x}))>0$. 
\end{itemize}
\end{definition}

\subsection{Information Efficiency of Primitive Recursive Functions} 
 We can construct a theory about the flow of information in computation. For  primitive recursive functions we follow \cite{Odi16}.  Given  $\log 0 = 0 $ the numbers $0$ and $1$ contains no information. 

\begin{itemize}
\item  The \emph{constant function} $z(n)=0$ carries no information $z^{-1}z(n)= \mathbb{N}$.  
\item The first application of the \emph{successor function} $s(n)=x+1$ is information conserving:  \[\delta(s(0))=  \log (0 + 1) - \log 0 = 0\] 
\item The \emph{successor function} expands information for values $>1$. By equation \ref{MEASINFGROW} we have:   
\[ I(s) = \log (x + 1) =  \Sigma_{n=1}^{\infty}(-1)^{n-1}\frac{x^n}{n} > \log x \]
Consequently: 
  \[\delta(s(x))= I(s(x)) - \log x - \log 1=  \log (x + 1) -\log x = \epsilon > 0\]
  Note that $\epsilon$ goes to zero as $x$ goes to $\infty$. 
\item The \emph{projection function} $P_{i,n}((x_1,x_2,\dots,x_n) = x_i$, which returns the i-th argument $x_i$,  is information discarding. Note that the combination of the index $i$ and the ordered set $(x_1,x_2,.., x_n)$ already specifies $x_i$ so: 
 \[\delta(P_{i,n}(x_1,x_2,.., x_n)=  I(x_i) - \log i - I(x_1,x_2,\dots, x_n) <  0\]  
\item \emph{Substitution.} If $g$ is a function of $m$ arguments, and each of $h_1,\dots,h_m$  is a function of $n$ arguments, then the function $f$:
\[f(x_1,\dots,x_n)=g(h_1(x_1,\dots,x_n),\dots,h_m(x_1,\dots,x_n))\]
is definable by composition from $g$ and $h_1,\dots,h_m$. We write $f=[g \circ h_1,\dots,h_m]$, and in the simple case where $m=1$  and $h_1$  is designated $h$, we write $f(x)=[g \circ h](x)$. Substitution is information neutral: 
\[\delta( f(x_1,\dots,x_n))=\delta(g(h_1(x_1,\dots,x_n),\dots,h_m(x_1,\dots,x_n))) =\]
\[I(f(x_1,\dots,x_n) - I(x_1,\dots,x_n) \]
Where $I(f(x_1,\dots,x_n)$ is dependent on $\delta(g)$ and $\delta(h)$. 
\item \emph{Primitive Recursion.} A function $f$ is definable by primitive recursion from $g$ and $h$ if $f(x,0)= g(x)$ and $f(x,s(y))=h(x,y,f(x,y))$.  Primitive recursion is information neutral: 
\[\delta(f(x,0))= \delta(g(x)) = I(g(x) - I(x)\]
which dependent on $ I(g(x)$ and 
\[\delta(f(x,s(y)))=\delta(h(x,y,f(x,y))) = I(h(x,y,f(x,y))) - I(x) - I(y)\]
which is dependent on $I(h(x,y,f(x,y)))$.
 \end{itemize}  
  
 Summarizing: the primitive recursive functions have one information expanding operation, \emph{counting}, one information discarding operation, \emph{choosing}, all the others are information neutral. 
 
 \subsection{Information Efficiency of Elementary Arithmetical Operations}\label{INFPRIMRECFUNC}
 
 The information efficiency of more complex operations is defined by a combination of counting and choosing. We analyse  the definition of addition by primitive recursion:  suppose $f(x) = P^1_1(x) = x$ and $g(x,y,z)= S(P_2^3(x,y,z)) = S(y)$. Then $h(0,x) = x$ and $h(S(y),x) = g(y,h(y,x),x) = S(h(y,x))$. The information efficiency is dependent on the balance between choosing and counting in the definition of $g$: 
  \[\delta(g(x,y,z))=  I(S(y)) - I(x,y,z) = (\Sigma_{n=1}^{\infty}(-1)^{n-1}\frac{y^n}{n}) - I(x,y,z) \]
  Note that the operation looses information and the information loss is asymmetrical. The commutativity and associativity of information efficiency for complex arithmetical operations are not trivial:   
  
\begin{theorem}[Information Efficiency of Elementary Arithmetical Operations]\label{BASICRULES}
\begin{itemize}  
\item \emph{Addition of different variables is information discarding}.  In the case of addition we know the total number of times the successor operation has been applied to both elements of the domain: for the number $c$ our input is restricted to the tuples of numbers that satisfy the equation $a + b =c $ $(a,b,c \in \mathbb{N})$. Addition is information discarding for numbers $>2$: 
 \begin{equation}\label{EA1}
 \delta(x + y)= \log (x + y) -\log x - \log y < 0 
 \end{equation}
 
\item \emph{Addition of the same variable has constant information}. It measures the reduction of information in the input of the function as a constant term:
\begin{equation}\label{EA2}
\forall (x) \delta(x + x)=  \log 2x -\log x  = \log 2 
\end{equation}

\item \emph{Commutativity for addition is information efficient:} 
\begin{equation}\label{COMMADD}
\delta (x + y) = \log (x + y) -\log x - \log y = \log (y + x) -\log y - \log x  = \delta (y + x)
\end{equation}

\item \emph{Associativity for addition is not information efficient:} 
\begin{equation}\label{ASSADD}
\begin{split}
\delta(x + (y + z)) = \log (x + (y + z) ) -\log x - \log (y + z) \neq \\\log ((x + y) + z ) -\log (x + y) - \log z =  \delta ((x + y) + z)
\end{split}
\end{equation}
  \item  \emph{Multiplication of different variables is information conserving.} In the case of multiplication  $a \times b =c $ $(a,b,c \in \mathbb{N})$ the set of tuples that satisfy this equation is much smaller then for addition and thus we can say that multiplication carries more information. If $a=b$ is prime (excluding the number $1$) then the equation even identifies the tuple.  Multiplication is information conserving for numbers $>2$: 
\begin{equation}\label{EA3}
 \delta(x \times y) = \log (x \times y) -\log x - \log y = 0 \end{equation} 
 
 \item \emph{Multiplication by the same variable is information expanding}. It measures the reduction of information in the input of the function as a logarithmic term:
\begin{equation}\label{EA4}
\forall (x) \delta(x \times x) = \log (x \times x) -\log x  = \log x > 0
\end{equation}

\item \emph{Commutative for multiplication is information efficient:} 
\begin{equation}\label{COMMUL}
\delta(x \times y) = \log (x \times y) -\log x - \log y =   \log (y \times x) -\log y - \log x = \delta(y \times x)
\end{equation}

\item \emph{Associativity for multiplication is information efficient:}  
\begin{equation}\label{ASSMUL}
\begin{split}
\delta(x \times (y \times z)) = \log (x \times (y \times z) ) -\log x - \log (y \times z) = \\ \log ((x \times y) + z ) -\log (x \times y) - \log z =  \delta ((x \times y) \times z)
\end{split}
\end{equation}
\end{itemize}

\end{theorem}

Estimating the information efficiency of elementary functions is not trivial. From an information efficiency point of view the elementary arithmetical functions are complex families of functions that describe computations with the same outcome, but with different computational histories. The non-associativity of addition specified in equation \ref{ASSADD} illustrates this. The associativity of information efficiency for multiplication is an `accident' caused by the information conservation of the operation, but as soon as we analyse it as addition of logarithms the same anomalies emerge. Summarizing: some arithmetical operations expand information, some have constant information and some discard information. The flow of information is determined by the succession of types of operations, and by the balance between the complexity of the operations and the number of variables.  These observations can be generalized to polynomial functions.

\subsection{Information Efficiency of  Polynomial Functions under Maximal Entropy } 

\begin{definition}[Polynomial Functions]\label{POLFUNC}
\begin{itemize}
\item A polynomial $p$ in the $k$ variables $x_1,x_2,\dots,x_k$ is a (finite) sum (by repeated application of the addition operation) of expressions of the type $c_iv_1v_2v_3 . . . v_j$ (or simply $c_0$) (formed by repeated application of the multiplication operation) where the coefficients $c_i$ are integers (positive or negative) and $v_i$ are variables. 
\item A polynomial is positive if all the coefficients are positive. 
\item A polynomial function is $f(x_1,x_2,\dots,x_k)=p$ where $p$ is a polynomial with $x_1,x_2,\dots,x_k$ as variables. 
\item The expression $p=0$ where $p$ is a polynomial is called a diophantine equation. 
\end{itemize}
\end{definition}

\begin{definition}[Information Efficiency for Polynomial Functions]\label{RANDEFPOLFUNC}
The information in a polynomial is measured as $I(p)=\log p$. The information efficiency of a polynomial function is $\delta(f(\overline{x}))= \log f(\overline{x}) - I(\overline{x})$  in line with definition \ref{EFFFUNCTION}. The information efficiency of a diophantine equation $p=0$ is $\delta(f(\overline{x}))= \log f(\overline{x}) - I(\overline{x}) < 0$, i.e. it discards all information internally. 
\end{definition}

\begin{definition} [Typical Sets] \label{TYPSET}
A set $s = \{x_,x_2,\dots ,x_k\}$ is \emph{typical} if it is random. For these sets we may estimate $I(s)= \Pi_{x_i in s}x_i$. In the limit we have an infinity of typical sets of any cardinality with equal amounts of information, see Appendix \ref{SETSINCOMP}.
\end{definition}

A central result is:  

\begin{theorem}[Conservation of Information for Polynomial  Functions under Maximal Entropy ]\label{LIMITRANDDEF} 
Suppose $f(\overline{x}))=p$ is a polynomial function of degree $m$ with $k$ variables, then in the limit for a dense segment of typical sets: 
\begin{itemize}
\item  $p$ has unbounded negative information efficiency (i.e. it discards information)  if  $k>m$. 
\item  $p$ has constant information efficiency (i.e. it conserves information) if  $k=m$.  
\item  $p$ has unbounded positive information efficiency (it creates compressible information) if  $k < m$. 
\end{itemize}
\end{theorem}

Proof: 
 By definition  \ref{EFFFUNCTION} the information efficiency is: 
\begin{equation}\label{EQ1}
\delta(f(\overline{x}))= \log f(\overline{x})) - I(\overline{x})
\end{equation}
Since $\{ x_1, x_2,\dots,x_k\}$ is typical the corresponding limit is: 
\begin{equation}\label{EQ2}
 \lim_{ x_1,x_2,\dots,x_k \rightarrow \aleph_0} \log f( x_1, x_2,\dots,x_k) - (\log x_1 + \log x_2,\dots,\log x_k) =
\end{equation}
\begin{equation}\label{EQ3}
\log  \lim_{ x_1,x_2,\dots,x_k \rightarrow \aleph_0} \frac{f( x_1, x_2,\dots,x_k)}{x_1,x_2, x_k} =
\end{equation}

\begin{equation}\label{EQ4}
\log  \lim_{ x_1,x_2,\dots,x_k \rightarrow \aleph_0}   \frac{
c_1x_1^{a_1}x_2^{b_1} . . . x_k^{k_1}
}{ x_1 x_2 \dots x_k}
+ 
\frac{
c_2x_1^{a_2}x_2^{b_2} . . . x_k^{k_2}
}{ x_1 x_2 \dots x_k}
+ \dots + 
\frac{
c_ix_1^{a_i}x_2^{b_i} . . . xk^{k_i}
}{x_1 x_2 \dots x_k} 
\end{equation}
Here $c_i$ is a positive or negative constant and $(a_i + b_i + c_i, . . . + k_i) \leq m$. : 
\begin{itemize}
\item Case  $m < k$. We evaluate expression \ref{EQ4}.  Each term  $\frac{ c_j x_1^{a_j} x_2^{b_j} x_3^{c_j} . . . n^{k_j} }{ x_1 x_2 \dots x_k}$ will be zero: \ref{EQ4} goes to $-\infty$.
\item Case  $m = k$. We evaluate expression \ref{EQ4}. Some terms $\frac{ c_j x_1^{a_j} x_2^{b_j} x_3^{c_j} . . . n^{k_j} }{ x_1 x_2 \dots x_k}$ will go to value $c_j$ under the condition $x_1=x_2=\dots=x_k$. All others will be zero.  We have to prove that for any information neighborhood $[I(x_i) , I(x_i) + \epsilon_1]$ there are `enough' typical sets with maximal entropy for which equation \ref{EQ4} goes to $c + \epsilon_2$ in the limit, i.e. the density of sets for which this holds is $>0$. This condition is satisfied by lemma \ref{ENTROPYLEMMA} in Appendix \ref{SETSINCOMP}. An alternative proof using transfinite information measurement is given in Appendix \ref{TRANSFININFMAN}. Equation  \ref{EQ4} is constant in the limit for a dense segment of typical sets if $k=m$. 

\item Case If $m >  k$. Under this constraint  \ref{EQ4} is information expanding if the information limit of the following polynomial is unbounded: 
\begin{equation}
c_1x_1^{a_1}x_2^{b_1} . . . x_k^{k_1}
+ 
c_2x_1^{a_2}x_2^{b_2} . . . x_k^{k_2}
+ \dots + 
c_ix_1^{a_i}x_2^{b_i} . . . x_k^{k_i}
\end{equation}
Which is the case for all polynomials (e.g. see Appendix \ref{POLUNBOUND}). 
\end{itemize}
$\Box$

\subsection{Information Efficiency of Diophantine Equations under Maximal Entropy}

The final step of this analysis is: 

\begin{lemma}[Conservation of Information of Diophantine Equations under Maximal Entropy ]\label{INFLIMDIOPH}
Let $p=0$ be a diophantine equation of degree $m$ with $k$ variables then, if $m > k$ it has no typical solutions in the limit. 
\end{lemma}
Proof: The expression $p=0$  implies that information efficiency of the polynomial $p$ equals that of the constant function: i.e. in the limit  $p$ annihilates all the information in its input variables: $\delta(p) < 0$. Now apply theorem  \ref{LIMITRANDDEF}. $\Box$

 The MRDP theorem states that a set of integers is Diophantine if and only if it is computably enumerable. So with lemma \ref{INFLIMDIOPH} we have a general theory about the flow of information in typical computable enumerable sets. Note that theorem \ref{LIMITRANDDEF} with lemma  \ref{INFLIMDIOPH} only hold for typical sets. In general it does not help us to say something about individual solutions for diophantine equation, but we can proive statments about the density of the sets of solutions.

 \section{Some Applications}
 
 \subsection{Density of Solutions for Fermat Equations}
 
 To show the power of  the maximal entropy approach of theorem \ref{LIMITRANDDEF} consider the following result.  A special case of  lemma \ref{INFLIMDIOPH} is the so-called last theorem of Fermat: 
 
\begin{lemma}[Fermat's Last Theorem]\label{FERMTHEM}
Diophantine equations of the form  
\begin{equation}\label{EQF}
x^n + y^n = z^n
\end{equation}
only have typical solutions for values of $n<3$. 
\end{lemma}
Proof: Special case of \ref{INFLIMDIOPH}. $\Box$

Note that this specific result only holds for typical value assignments. The density of solutions for equations with $n>2$ is zero in the limit. The constraint imposed by lemma \ref{FERMTHEM} is much weaker than the one implied by Faltings' theorem by which the amount of solutions finite. A general proof of this theorem is given in \cite{WIL95}. Theorem \ref{LIMITRANDDEF} does not help us to prove statements about the individual existence of solutions for diophantine equations. 

\subsection{Perfect Information Efficiency of Cantor Pairing Function}

In some cases the density estimates of theorem \ref{LIMITRANDDEF} with lemma  \ref{INFLIMDIOPH} are sufficient to prove a result. We can deploy it in a technique that one could call \emph{proof by countability}: Since by theorem \ref{INCOMPRESSIBILITY}  the set of natural numbers is incompressible any polynomial function that defines a bijection between a countable set and the set of natural numbers must be information efficient in the limit: i.e. is should have constant information efficiency for all typical sets.

The set of natural numbers $\mathbb{N}$ can be mapped to its product set by the so-called Cantor pairing function $\pi^{(2)}: \mathbb{N} \times \mathbb{N} \rightarrow \mathbb{N}$ that defines a two-way polynomial time computable bijection:

\begin{equation}\label{CANTORPAIRING}
\pi^{(2)}(x,y) := \frac{1}{2}(x + y)(x + y + 1)+y
\end{equation}
 
The Fueter - P\'{o}lya theorem \cite{FP23} states that Cantor pairing is the only possible quadratic pairing function. Since it defines a bijection with the natural numbers it must in the limit be information efficient, which is consistent with lemma \ref{LIMITRANDDEF} that predicts a constant information efficiency. Note that for sets of tuples $(0,y)$ and $(x,0)$ equation \ref{CANTORPAIRING} collapses into an information expanding zero-density function by theorem \ref{LIMITRANDDEF}. The same theorem gives us a proof for a conjecture by Fueter and P\'{o}lya  \cite{FP23}: 

\begin{lemma}[Fueter-P\'{o}lya Generalisation]\label{FUTPOLGEN}
There is no polynomial function $f: \mathbb{N} \times \mathbb{N} \rightarrow \mathbb{N}$ of degree $>2$ that defines a bijective mapping. 
\end{lemma}
Proof: immediate consequence of theorem \ref{LIMITRANDDEF}. $f$ would have degree $>2$ with $2$ variables and thus will be information expanding on the dense set of all typical value assignments, but this contradicts the fact that it is a bijection to $\mathbb{N}$. $\Box$

 Note that there exist many other polynomial time computable bijective mappings between $\mathbb{N} \times \mathbb{N}$ and $\mathbb{N}$ (e.g. Szudzik pairing)\footnote{See http://szudzik.com/ElegantPairing.pdf, retrieved January 2016.} and $\mathbb{N}^k$ and $\mathbb{N}$. By theorem \ref{LIMITRANDDEF} none of these mappings can be information expanding or discarding in the limit. 

\subsubsection{Using the Cantor Function to Define Countable Information Partitions of $\mathbb{N}$}

 By lemma  \ref{FUTPOLGEN} we have an information conserving  mapping from $\mathbb{N} \times \mathbb{N}$ to $\mathbb{N}$. We can interpret $\mathbb{N} \times \mathbb{N}$ as an information set that allows us to split any natural number into two new numbers without information loss. We will call the object defined by the Cantor pairing function a \emph{Cantor Partition on $\mathbb{N}$} with the notation $\mathcal{G}^2(\mathbb{N})$. The Cantor partition $\mathcal{G}^2(\mathbb{N})$ is a special infinite set that has three types of index functions: the \emph{horizontal row index function},  a \emph{vertical column index function}, and a \emph{diagonal 'counting' index function} defined by the Cantor function. 

Any function computing a bijection between the Cantor grid and the natural numbers must compute values whose information content is asymptotically close to that of the values computed by the Cantor pairing function. This is the motivation for calling a Cantor Partition an \emph{information partition}: it objectively matches the amount of information of elements of $\mathbb{N}$ with elements of $\mathbb{N} \times \mathbb{N}$. We can use the object $\mathcal{G}^2(\mathbb{N})$ as a measuring device:

\begin{lemma}[Countable Information Partitions of $\mathbb{N}$] \label{INFCOUNTPARTN}
Let $S$ be a countable infinite set partitioned in to an infinite number of infinite subsets $S_i$, then: 
\[\forall (s_{i,j} \in S_i \subseteq S) I(s_{i,j}|S) = \log \pi^{(2)}(i,j)\]
\end{lemma}
Proof: $i$ is the index of the set, $j$ is the index of the element in the set. By lemma  \ref{LIMITRANDDEF} the function $\pi^{(2)}$ has constant information efficiency, i.e. it assigns in the limit  $s_{i,j}$ exactly the value that is consistent with the information in $i$ and $j$. $\Box$

\subsection{Measuring Information in Finite Sets of Numbers}\label{INFINNUMBERS}

Let $\mathfrak{P}(\mathbb{N})$ be the set of finite subsets of $\mathbb{N}$. This has to be distinguished from $\mathcal{P}(\mathbb{N})$ which also contains all the infinite subsets of $\mathbb{N}$. $\mathcal{P}(\mathbb{N})$ is uncountable, whereas $\mathfrak{P}(\mathbb{N})$ can be counted. Proofs of the countability of $\mathfrak{P}(\mathbb{N})$ rely on the axiom of choice to distribute set  $\mathfrak{P}(\mathbb{N})$ in to partitions with the same cardinality.  We show that there is an algorithm that defines such a  polynomial time computable bijection between  $\mathfrak{P}(\mathbb{N})$ and $\mathbb{N} \times \mathbb{N}$. The partitions $S_k$ are described by combinatorial number systems of degree $k$. The function $\sigma_k:\mathbb{N}^k \rightarrow \mathbb{N}$ defines for each element $s = (s_k,\dots,s_2,s_1) \in \mathbb{N}^k$  with the strict ordering $s_k > \dots s_2 > s_1 \geq 0$ its index in a $k$-dimensional combinatorial number system as:

\begin{equation}
\sigma_k(s)= {s_k \choose k} + \dots + {s_2 \choose 2} + {s_1 \choose 1}
\end{equation}

The function $\sigma_k$ defines for each set $s$ its index in the lexicographic ordering of all sets of numbers with the same cardinality $k$. The correspondence does not depend on the size $n$ of the set that the $k$-combinations are taken from, so it can be interpreted as a map from $\mathbb{N}$ to the $k$-combinations taken from $\mathbb{N}$. The correspondence is a polynomial time computable bijection. A set $s \in \mathfrak{P}(\mathbb{N})$ has a one to one recursive index $\sigma(s)$ in a $|s|$-dimensional combinatorial number system, which gives:  

\begin{lemma}[Information in Sets of Numbers]\label{Information in Sets of Numbers}
If $s  \in \mathfrak{P}(\mathbb{N} ))$ is a finite set of numbers then the information in $s$ is: 
\[I(s) =  \log \pi^{(2)}((|s|),\sigma_{|s|}(s))\]
\end{lemma}
Proof: both $\pi^{2}$ and $\sigma$ are computable bijections. $\Box$. 

In this case, for each number $\pi^{(2)}(k_1,k_2)$ the number $k_2$ is the index of a partition of all sets with cardinality $k_1$. We will call the resulting information grid a \emph{Cantor partition  of $ \mathfrak{P}(\mathbb{N})$ ordered on cardinality} with the notation $\mathcal{G}^2_{card}( \mathfrak{P}(\mathbb{N}))$. The double occurrence of the term $|s|$ in the expression $\sigma_{|s|}(s))$ shows the `inner working' of the axiom of choice: it only serves as a meta-variable to select a formula of the corresponding arity. The Cantor partition $\mathcal{G}^2_{card}( \mathfrak{P}(\mathbb{N}))$ is a special infinite set that has three types of index functions for elements $s \in \mathfrak{P}(\mathbb{N})$ with $|s|=k$.  In the limit we'll have asymtotic information balance between: 
 \begin{itemize}
\item  the \emph{horizontal row index function} with information efficiency: \[\delta(|s|) = \log k  - I(s)\]  
\item   the  \emph{vertical column index function}. By theorem \ref{LIMITRANDDEF} it has information efficiency: \[\delta (\sigma_k(s))= \log  \sigma_k(s) - I(s) = c \] .
\item the \emph{diagonal 'counting' index function} defined by the Cantor function, which by theorem \ref{LIMITRANDDEF} has information efficicency: \[\delta( \pi^{(2)}((|s|),\sigma_{|s|}(s))) = c\] 
\end{itemize}

We have the following balance condition: 

\begin{lemma}[Information Balance on Cardinality]\label{INFBALCARD}
\[I(s) = \log |s|  +  \log \sigma_{|s|}(s) + c\]
\end{lemma}
Proof: By definition  $\pi^{(2)}(|s|,\sigma_{|s|}(s))$ has constant information efficiency, so: 

\[\delta (\pi^{(2)}(|s|,\sigma_{|s|}(s))) = \]\[ \log \pi^{(2)}(|s|,\sigma_{|s|}(s) - \log |s|  -  \log \sigma_{|s|}(s) =  I(s) - \log |s|  -  \log \sigma_{|s|}(s)= c \]
$\Box$ 

This balance function can be generalized to the Law of Conservation of Information  \ref{INFCONS} given in the summary of the proof at the beginning of the paper.

\subsection{Optimal Representation of Dimensional Information and Directed Graphs}
A nice consequence of this analysis is that we have an optimal representation of sets of numbers with dimension $k$ and dimensionless objects such as directed graphs. Such a representation is useful if we want to study the compressibility these objects.~\cite{BLO16} For sets with dimension $k$ we can use the corresponding combinatorial number system. Suppose we have graph $G$ of size $n$ is a tuple $(N, L)$ containing a set of nodes $N$ and a set of links $L$, where $N$ is the set of the first $n$ natural numbers. $L$ contains pairs of elements from $N$. Let $NG$ be the nodeset of $G$ and $LG$ be its linkset. If a graph $G$ is directed, the pairs in $LG$ are ordered and correspond to a set of points on the Cantor grid  $\mathcal{G}^2(\mathbb{N})$ that, using equation \ref{CANTORPAIRING}, can be represented as a set of natural numbers $s$.

\subsection{Enumerating Finite Sets by their Sums}
The function $I(s) =  \log \pi^{(2)}((|s|),\sigma_{|s|}(s))$ enumerates the set of all finite sets of numbers on their cardinality: it gives us the $n$-th set with cardinality $k$. By the Fueter and P\'{o}lya theorem and its generalisation lemma \ref{FUTPOLGEN} we know that there are no other polynomial functions that compute a similar mapping: i.e. the ordering on cardinalities, and rank in a combinatorial number system defines in the limit the only existing bijection between $\mathfrak{P}(\mathbb{N})$ and $\mathbb{N}$. We investigate this insight in the context of addition. 

Let $S_n = \{x \in \mathfrak{P}(\mathbb{N})| \wedge \Sigma s = n\}$ be the set of subsets of finite numbers that  add up to $n$. If we define $p(n) = |S_n|$. The number of these sets is estimated by the Hardy-Ramajunan asymptotic partition formula:  $p(n) \sim \frac{1}{4 n \sqrt{3}} e ^{\pi \sqrt{2n/3}}$. This means that for each $n$, the set $S_n$ is large but finite. Define $\Sigma s = \Sigma_{s_i \in s} s_i$ as the addition function for sets. From equation  \ref{ASSADD} in section \ref{INFPRIMRECFUNC} we know that associativity is not information efficient for addition. As a consequence: : 

\begin{lemma}[Information Efficiency Fluctuations for Addition are Unbounded]\label{RANDEFGENADD}

There is no constant $c$ such that for any two sets $s$ and $r$ of the same cardinality $k$ we have that  $\delta(\Sigma_{s_i \in s} s_i)  - \delta(\Sigma_{r_i \in r} r_i)| < c$. There are no functions of arity $k$ that are more efficient. 
\end{lemma}
Proof: 
We prove the base case $k=3$. Suppose $f(x,y,z) = x + y + z$ then for any constant $c$ there are positive numbers such that:
\[ |\delta ((x + y) + z) -  \delta(x + (y + z)) | < c\]
\[\delta ((x + y) + z) = \log ((x + y) + z ) -\log (x + y) - \log z \]
\[\delta(x + (y + z)) = \log (x + (y + z) ) -\log x - \log (y + z) \] 
Consider the expression: 
 \[|(\log ((x + y) + z ) -\log (x + y) - \log z) - (\log (x + (y + z) ) -\log x - \log (y + z))|<  c\]
For any constant we can choose a combination of $z$ and $x$ that makes this inequality invalid. 
$\Box$

The rationale behind these differences is clear. Suppose $\{x,y,z\}$ has maximal entropy, and $m=x+z$. Then the entropy of the set $\{m, z\}$ is relatively lower: the expression $((x + y) + z )$ is associated with first picking one element from a high entropy set, and then picking one element from a low entropy set. Expression $f(x,y,z)$ abstracts from these histories and is associated with picking three elements from a set with replacement. This proof can be easily generalized for addition over any bounded number of variables. Suppose $f(x,y,(z_1 + \dots  + z_k)) = x + y + (z_1 + \dots  + z_k)$ then for large enough numbers: $\delta ((x + y) + (z_1 + \dots  + z_k)) \neq \delta(x + (y + (z_1 + \dots  + z_k)))$. The information efficiency of the addition operation is dependent on the succession of choices of variables during the computation and the entropy of the underlying set and is relatively independent of the actual value of the sum. As a consequence sets cannot be enumerated on their sums:  

\begin{theorem}[No Polynomial Function Enumerates Equal Sum Partitions]\label{HURRAY}
There is no polynomial function that enumerates the set $\mathfrak{P}(\mathbb{N})$ on its equal sum partitions.
\end{theorem}
Proof: Suppose $\theta$ is the function that enumerates sets with the same sum, and $\eta$ is the set that maps the pair of values bijectively to $\mathbb{N}$: 
\[\eta(\Sigma s,\theta_{\Sigma s}(s))\]

In the limit we'll have asymtotic information balance between the horizontal row index function, the vertical column index function and the information in the set: 
 \begin{itemize}
\item  the \emph{horizontal row index function} with information efficiency: $\delta(\Sigma_s)$. Here the term $\Sigma_s$ designates an infinite set of functions, for which by lemma \ref{RANDEFGENADD} the information  efficiency fluctuates unboundedly. 
\item   the  \emph{vertical column index function} defined by the unknown function $\theta_{\Sigma_{s}}(s)$,
\item the \emph{diagonal 'counting' index function} defined by the unknown function  $\eta(\Sigma_{s},\theta_{\Sigma_{s}}(s))$. 
\end{itemize}

Although we do not know function $\theta$ and $\eta$ they must satisfy the following conditions. By definition  $\eta(\Sigma_{s},\theta_{\Sigma_{s}}(s))$ has constant information efficiency, so in the limit we have: 
\[I(s) =  \log \eta(\Sigma_{s},\theta_{\Sigma_{s}}(s)) + c\]
Which defines a balance condition analagous to the one in lemma \ref{INFBALCARD}: 
\begin{equation}\label{BALANCESUM}
I(s) = \log \Sigma s  +  \log \theta_{\Sigma_{s}}(s)+ c
\end{equation}
Note that with this step the function $\eta$ is eliminated from our analysis. We only have one unknown: the function $\theta$. It is clear that equation \ref{BALANCESUM} can be satisfied for sets $s$ of cardinality $|s| = 2$: 
\[I(\{x,y\}) = \log (x+y) +  \log \theta_{2}(\{x,y\}) + c\]
But already for $|s| = 3$ there is no adequate interpretation: 
\[I(\{x,y,z\}) = \log( (x+y)+z) +  \log \theta_{3}(\{x,y,z\}) + c\]
Equation \ref{BALANCESUM} has no general solution. In order to satisfy equation \ref{BALANCESUM} for all finite sets we need:
\begin{itemize} 
\item \emph{An Infinite number of versions of $\theta_k$ for each natural number $k$.} For each set with $k$ elements $\theta$ is a function with arity $k$, it can only have constant information efficiency if there is an unlimited number of versions of $\theta_k$ for each finite number $k$. But the $\log \Sigma_s$ function does not contain that information, so selection of the right $\theta_k$ is not possible.
\item \emph{If $k$ goes to $\infty$ an unbounded number of different versions of $\theta_k$ for fixed $k$.} Even if we select a $\theta_k$ for sets of a specific cardinality, the fact that $\Sigma s$ has no stable definition prevents a satisfying assignment for equation \ref{BALANCESUM}. According to lemma \ref{RANDEFGENADD} the information  efficiency of $\Sigma s$ as an infinite collection of functions can vary unboundedly.   
\end{itemize} 
$\Box$ 
 
\section{Discussion}

\subsection{Some consequences of lemma \ref{HURRAY}}

By theorem \ref{Information in Sets of Numbers} we know that we can count the set of finite sets $\mathfrak{P}(\mathbb{N})$ ordered on cardinality.  By theorem \ref{HURRAY} we know that we cannot count the same sets ordered by sum. There is a function that gives us the $n$-th set of size $k$ but there is no polynomial function that gives us the $n$-th set that adds up to $k$. We can enumerate sets with equal sums but only via the ordering on cardinality (See Appendix \ref{COUNTSETSUM} for an example).  The consequence is that when we want to search for subsets that add up to $k$, the only ordering we can use is the one on cardinality. This gives a strategy to prove that there is no general short solution for the so-called subset sum problem (See appendix \ref{SUBSETSUM} for a discussion). Another consequence of theorem \ref{HURRAY} is 

\begin{lemma}[No Polynomial Function Enumerates Equal Product  Partitions]\label{HURRAY2}
There is no polynomial function that enumerates the set $\mathfrak{P}(\mathbb{N})$ on its equal product partitions.
\end{lemma}
Proof: Analogous to the proof for theorem \ref{HURRAY}: replace $\Sigma s = \Sigma_{s_i \in s} s_i$ with  $\Sigma s = \Sigma_{s_i \in s} \log s_i$. $\Box$

This implies that when we want to search for subsets that multiply to $k$, the only ordering we can use is the one on cardinality. This gives a strategy to prove that there is no general solution for the factorization problem.

\subsection{Relation with other Information Measures}

These results give an indication about the way this computational theory of information is related to two other concepts of information: Shannon information and Kolmogorov complexity.
\subsubsection{Kolmogorov Complexity}

 Theorem \ref{INCOMPRESSIBILITY}  is equivalent to the counting argument used in Kolmogorov complexity \cite{LiVi08} to prove that most strings are incompressible and thus random. It is an asymptotic measure, but not in the sense that one has to take a constant $O(1)$ into account as compensation for the selection of an arbitrary Turing machine. Mutatis mutandis one might say that most natural numbers are random. This argument holds for general functions and does, unlike Kolmogorov theory, not depend on some theory of computation. In any formal system strong enough to represent the set of natural numbers the overwhelming majority of numbers will be random. The theorem proves that most natural numbers are incompressible in the limit. It does not define the concept of randomness in itself, but as soon as one plugs in a definition of compressibility with unbounded compression, the theorem starts to work for us. An example would be: A number $n$ is compressible if there is a function $f: \mathbb{N} \rightarrow \mathbb{N}$ such that $\log f(n) < \log n - \log \log \log n$, with the unbounded compression function $\log  \log n$. Given the fact that there is an unlimited number of unbounded compression functions in principle there is an unlimited number of different definitions of the concept of randomness. 

\subsubsection{Shannon Information}

There is also a relation between compression functions and Shannon information.~\cite{SHA48} By definition we have for a set $A \subseteq \mathbb{N}$ with compression  function $c_{A}$: 
\[\frac{c_{A}(n)}{n}  \times \frac{n - c_{A}(n)}{n} = 1\]
Here $P(x \in A) = \frac{c_{A}(n)}{n}$ is an estimate of the probability that x is an element of $A$ based on the first $n$ observations and $P(x \in \neg A) =  \frac{n - c_{A}(n)}{n}$ an estimate of the probability that $x$ is not an element of $A$ based on the first $n$ observations.  We can estimate the Shannon entropy of the set $A \in \mathbb{N}$ as: 
\[H(A \subseteq \mathbb{N}) =  -(P(x \in A) \log  P(x \in A) + P(x \in \neg A) \log P(x \in \neg A)) \]
Under the central limit theorem this condition is satisfied and Shannon Information is defined in the limit, but there are many data sets for which density is not defined in the limit and consequently the application of Shannon theory makes no sense. 

\section{Acknowledgements}
I thank Steven de Rooij, Peter Bloem, Frank van Harmelen, Amos Golan, Erik Schultes, Peter van Emde Boas, Rini Adriaans and my fellows at the Info-metrics Institute for many inspiring discussions. I thank the University of Amsterdam and especially Cees de Laat for allowing me the leeway to pursue my research interests. This research was partly supported by the Info-Metrics Institute of the American University in Washington, the Commit project of the Dutch science foundation NWO, the Netherlands eScience center, the ILLC and the IvI of the University of Amsterdam and a Templeton Foundation’s Science and Significance of Complexity Grant supporting The Atlas of Complexity Project.

\bibliographystyle{plain}
\bibliography{adriaans}

\newpage

\section{Appendix: The Subset Sum Problem and  Search by Partial Description}\label{SUBSETSUM}

\emph{And how will you enquire, Socrates, into that which you do not know? What will you put forth as the subject of enquiry? And if you find what you want, how will you ever know that this is the thing which you did not know?} This question, known as Meno's paradox, has interested philosophers for more than two millenia. In the past decennia mathematicians have been pondering about a related question: suppose it \emph{would} be easy to check whether I have found what I'm looking for, how hard can it be to find such an object? In mathematics and computer science there seems to be a considerable class of decision problems that cannot be solved constructively in polynomial time ($t(x)=x^c$, where $c$ is a constant and $x$ is the length of the input) , but only through systematic search of a large part of the solution space, which might take exponential time ($t(x)=c^x$). This difference roughly coincides with the separation of problems that are computationally feasible from those that are not.

The issue of the existence of such problems has been framed as the possible equivalence of the class $P$ of decision problems that can be solved in time polynomial to the input to the class $NP$ of problems for which the solution can be checked in time polynomial to the input. A well-known example in the class NP is the so-called subset sum problem: given a finite set of natural numbers $S$, is there a subset that sums up to some number $k$? It is clear that when someone proposes a solution $X \subseteq S$ to this problem we can easily check wether the elements of $X$ add up to $k$, but we might have to check almost all subsets of $S$ in order to find such a solution ourselves.

\subsection{Search and descriptive complexity}
In the past decennia a plethora of complexity classes  and their interrelations have been identified.~\footnote{See e.g. https://en.wikipedia.org/wiki/List\_of\_complexity\_classes, 
retrieved May 25 2016.} In this paper we will not deal with complexity classes as such, but with the philosophical and mathematical issues that help us to understand why they exist and what separates them. A number of questions are relevant in this context. Does computation create new information? Is there a difference between construction and systematic search? Are there uniquely identifying descriptions that do not contain all information about the object they refer to? 

Since Frege most mathematicians seem to believe that the answer to the last question is positive.~\cite{FREGEB} The descriptions "The morning star" and "The evening star" are associated with \emph{procedures} to identify the planet Venus, but they do not give access to all information about the object itself. If this were so the discovery that the evening star is in fact also the morning star would be uninformative. Yet in constructive mathematics such a position leads to counterintuitive results. In Kolmogorov complexity for instance we define the information content of an object as the shortest program that produces the object in finite time on empty input. Now take the description "The first number that violates the Goldbach conjecture" and you have an apparent paradox. If this number exists it can be found by a very small program and does not contain much information. How hard can it be to find a simple object? Yet, if it does not exist we can look forever, without finding it. There is a similar paradox in the subset sum problem. If the set $S$ does contain a subset $X$ that adds up to $k$ then this set does not contain much information conditional to the definition of $S$: it is simply: "The first subset of $S$ that adds up to $k$ that we find when we search systematically". 

In order to avoid such problems Russell proposed to interpret unique descriptions existentially~\cite{RUSS05}: A sentence like "The king of France is bold" would have the logical structure $\exists (x) (KF(x) \wedge \forall (y)(KF(y) \rightarrow x=y) \wedge B(x))$, but this theory does not help us to analyse decision problems that deal with existence. Suppose the predicate $L$ is true of $x$ if I'm looking for $x$, then the logical structure of the phrase "I'm looking for the king of France" would be $\exists (x) (KF(x) \wedge \forall (y)(KF(y) \rightarrow x=y) \wedge L(x))$, i.e. if the king of France does not exist it cannot be true that I am looking for him. The implication of Russell's theory for constructive mathematics seems to be that we can only search for objects on the basis of partial descriptions if we are sure to find them in a finite time, which is unsatisfactory. This is specifically so in the context of the subset sum problem, where it is very well possible that we are searching in a bounded domain for a set that does not exist. It seems that the Meno paradox is not about existence but about information. If there is a king of France, the predicate $KF$ gives only partial information about the real individual. We get more information when we finally meet the king in person and we can check, amongst many other things, whether he is bald or not.  

This analysis prima facie appears to force us to accept that in mathematics there are simple descriptions that allow us to identify complex objects by means of systematic search. When we look for the object we have only little information about it, when we finally find it our information increases to the set of full facts about the object searched. This is at variance with the observation that deterministic computing can never create new information, which is analyzed in depth in \cite{AB2011}. In a deterministic world it is the case that if  \texttt{program(input)=output} then $I(\texttt{output}) \leq I(\texttt{program}) + I(\texttt{input})$. The essence of information is uncertainty and a message that occurs with probability "1" contains no information. 

\subsection{The $\mu$-operator for unbounded search}

There is a subtle difference between systematic search and deterministic construction that is blurred in our current definitions of what computing is. If one considers the three fundamental equivalent theories of computation, Turing machines, $\lambda$-calculus and recursion theory, only the latter defines a clear distinction between construction and search, in terms of the difference between primitive recursive functions and $\mu$-recursive functions. The  set of primitive recursive functions consists of: the  constant function, the successor function, the projection function, composition and primitive recursion. With these we can define everyday mathematical functions like addition, subtraction, multiplication, division, exponentiation etc. In order to get full Turing equivalence one must add the $\mu$-operator. In the world of Turing machines this device coincides with infinite loops associated with undefined variables. It is defined as follows in \cite{Odi16}:

 For every 2-place function $f(x,y)$ one can define a new function, $g(x) = \mu y[f(x,y)=0 ] $, where $g(x)$ returns the smallest number y such that $f(x,y) = 0.$  Defined in this way $\mu$ is a partial function. It becomes a total function when two conditions are added: 
\begin{enumerate}
\item there actually exists at least one $z$ such that $f(x,z) = 0$ (this is in fact Russell's solution in his theory of descriptions) and 
\item for every $y\prime \leq y$, the value $f(x,y\prime)$ exists and is positive (i.e. $f$ is a total function). 
\end{enumerate}
One way to think about $\mu$ is in terms of an operator that tries to compute in succession all the values $f(x,0)$, $f(x,1)$, $f(x,2)$, ... until for some $m$ $f(x,m)$ returns $0$, in which case such an $m$ is returned. In this interpretation, if $m$ is the first value for which $f(x,m) = 0 $ and thus $g(x) = m$, the expression  $\mu y[f(x,y)=0 ]$ is associated with a routine that performs exactly  $m$ successive test computations of the form $f(x,y)$ before finding $m$. Since the $\mu$-operator is unbounded $m$ can have any value. Consequently the descriptive complexity of the operator in the context of primitive recursive functions is unbounded. We are now in a position to specify the difference between construction and search in more detail: 

\begin{itemize}
\item \emph{Descriptive power versus computation time.} The more powerful your theory of computing is, the more efficient your descriptions of objects become. Primitive recursive functions are provably weaker than $\mu$-recursive functions so a description of an object in the former formalism will possibly be longer, say $k$ bits. The shorter descriptions in the context of $\mu$-recursive functions do not imply reduction in time necessary to compute the object: we still have to do a sequence $2^k$ of computations which takes exponential time. This explains the Meno paradox in computation. Search is fundamentally different from construction. The introduction of unbounded search operations in a computational system causes the descriptive complexity of the objects to decrease while the  time to compute those objects increases with the representational efficiency gained. 

\item \emph{General function schemes versus particular function names.} The name $g$ does not refer to a function but to a function-scheme. The $x$ in the expression $g(x)$ is not an argument of a function but the index of a function name $f_x(y) \Leftrightarrow f(x,y)$. We can interpret the $\mu$-operator as a meta-operator that has access to an infinite number of primitive recursive functions. In this interpretation there is no such thing as a general search routine.  Each search function is specific: searching for your glasses is different from searching for your wallet, even when you look for them in the same places. 
\end{itemize}

\subsection{The subset sum problem}

An informal specification of the subset sum problem in this context would be: 
\[g_k(S) = \mbox{The first subset $x$ of $S$ that adds up to $k$}\]

Based on the analysis above we can make some observations: 

\begin{itemize}
\item the condition that such a subset must exist (Russell's condition) can be dropped, because the domain is bounded. If we interpret the $\mu$-operator in the strict sense as a total functions we cannot use it here. We might be looking for something that does not exist.  
\item The function implies that the poset of subsets of $S$ of can be enumerated constructively, not only by virtue of the fact that $S$ is finite. 
\item The function that checks whether a set adds up to $k$ is primitive recursive and polynomial. 
\item The condition that the checking function must be total is fulfilled. 
\item The expression $g_k$ is the name of a function scheme and $g_k(S)$ is the name of an individual function.  
\end{itemize}
We formulate the following conjecture: 
\begin{conjecture}
Each instance $S$ of the subset sum problem can be solved by a partial general search routine $g_k$ working on a bounded domain in exponential time. This implies that there is an infinite number of particular partial functions of the form $g_k(S)$ that each solve a different variant of the problem. There is not one general function with a finite definition that solves all particular problems, so, a fortiori, there is no general primitive recursive function that solves all the instances of the problem in polynomial time. 
\end{conjecture}

A way to prove such a conjecture would be to show that a general solution for the subset sum problem would involve a function that reduces a set with an infinite amount of information to a set with a finite amount of information. In the main part of this paper such a statement is proved: there are sets that have infinite descriptive complexity that can be defined by sequential execution of an infinite number of functions. Such sets cannot be described by a single function with finite descriptive complexity, yet instances of decision problems defined on such sets can be solved by Turing equivalent systems in exponential time. 

The proof technique is based on the incompressibility of the set of natural numbers. The set of finite sets natural numbers is \emph{rigid}: the set can be ordered constructively on cardinality. There are information discarding recursive functions defined on finite sets of numbers that are \emph{elastic}: they retain in the limit considerably more information than the cardinality of the sets. Such sets contain infinite information and thus cannot be described by one specific function. Search in such sets is non-constructive: i.e. decision problems about the existence of sets defined in terms of elastic functions is possible in exponential- but not in polynomial time. Addition of sets of numbers is an elastic primitive recursive function that has non-zero density in the limit. This implies that there is no general polynomial time function that solves the subset sum problem. This separates the class $P$ form the class $NP$.  

\section{Appendix: Transfinite Information Measurement} \label{TRANSFININFMAN}

We have observed that $\lim_{x \rightarrow \infty} I(x+1) - I(x) = 0$. In the limit the production of information by counting processes stops as specified by the derivative $\frac{d}{dx} \ln x = \frac{1}{x}$. This suggests that $I(x)$ has a definite value in the limit. We define: 
\begin{definition}[Small Infinity]\label{SMALLINF}
 \[\lim_{x \rightarrow \aleph_0}  - \log \frac{1}{x} = \aleph_{-1}\]
\end{definition}
For the moment $\aleph_{-1}$ is a purely administrative symbol without any interpretation. Equation \ref{SMALLINF} just allows us to substitute $ - \log \frac{1}{x}$ by $\aleph_{-1}$ in our derivations and vice versa. We investigate the basic rules of arithmetic for these substitutions. We can use a generalized Puiseux series\footnote{See Wolfram-Alfa: \tt{series log(x+1) at infty.}} to estimate the value of $\aleph_{-1}$. We define 
\[\tau_x = \frac{1}{x}-\frac{1}{2 x^2}+\frac{1}{3 x^3}-\frac{1}{4 x^4}+\frac{1}{5 x^5}-\frac{1}{6 x^6}+O((\frac{1}{x})^7)\]
The generalized Puiseux series for $x+1$ at $\infty$ is: 

\begin{equation}\label{GENPUIS}
\lim_{x \rightarrow \aleph_0}  \log ( x + 1) =  -\log(\frac{1}{x})+\tau_x =  \aleph_{-1}   + 0 = \aleph_{-1} \end{equation}

This implies that $\aleph_{-1}$ is  a definite entity with specific computational rules:

 $\aleph_{-1}$ absorbs addition by the fact that $\frac{1}{\aleph_0}= (\frac{1}{\aleph_0})^2$: 
 \begin{equation}\begin{split}\label{LIMAL1} 
\aleph_{-1} + \aleph_{-1} & =  \lim_{x \rightarrow \aleph_0}  \log ( x + 1) +   \lim_{x \rightarrow \aleph_0} \log ( x + 1) \\
  & =  -\log(\frac{1}{x})   +  \tau_x   + \log(\frac{1}{x})   -  \tau_x  \\
  & = -2\log(\frac{1}{x}) = -\log((\frac{1}{x})^2) = \log(\frac{1}{x}) \\ 
  &  = \aleph_{-1}
\end{split}\end{equation}
Equation \ref{LIMAL1} implies that also $\aleph_{-1} -  \aleph_{-1} = \aleph_{-1}$ and that for any two finite positive or negative variables $a$ and $b$ 
 \begin{equation}\begin{split}\label{LIMAL2}
 \aleph_{-1} \times a = \aleph_{-1} \times b
\end{split}\end{equation}
By contrast $\aleph_{-1}$ preserves multiplication and division: 
  \begin{equation}\begin{split}\label{LIMAL3}
 \aleph_{-1} \times  \aleph_{-1}&  =   \lim_{x \rightarrow \aleph_0}  \log ( x + 1) \times   \lim_{x \rightarrow \aleph_0} \log ( x + 1)\\ & =  (-\log(\frac{1}{x})   +  \tau_x )  \times  (-\log(\frac{1}{x})   +  \tau_x ) \\ 
 & =  -\log(\frac{1}{x})  \times  -\log(\frac{1}{x}) =  (-\log(\frac{1}{x}))^2 \\ & = (\aleph_{-1})^2 
\end{split}\end{equation}
 \begin{equation}\begin{split}\label{LIMAL4}
\frac{\aleph_{-1}}{  \aleph_{-1}} =  \frac{\lim_{x \rightarrow \aleph_0}  \log ( x + 1)} { \lim_{x \rightarrow \aleph_0} \log ( x + 1)}
 =  \frac{-\log(\frac{1}{x})   +  \tau_x }{-\log(\frac{1}{x})   +  \tau_x}  = 1 
\end{split}\end{equation}

 $\aleph_{-1}$ absorbs addition by terms of different degree. If $k < m$ then: 
 \begin{equation}\begin{split}\label{LIMAL5}
 (\aleph_{-1})^m - (\aleph_{-1})^k = (\aleph_{-1})^m  + (\aleph_{-1})^k = (\aleph_{-1})^m
\end{split}\end{equation}
Applying rule \ref{LIMAL3} en \ref{LIMAL4} we can devide equation \ref{LIMAL5} by $ (\aleph_{-1})^k$ which gives:
\[ (\aleph_{-1})^{m-k}  - 1 = (\aleph_{-1})^{m-k}  + 1 = (\aleph_{-1})^{m -k}\]

From the work of Cantor we know that in  the limit powersets are fundamentally bigger than their generating sets: 

\begin{equation}\label{LIM1}
|\mathbb{N}| = \aleph_0
\end{equation}

\begin{equation}\label{LIM2}
|\mathcal{P}(\mathbb{N}) |= 2^{\aleph_0} = \aleph_1
\end{equation}
We also know that $\aleph_0$ absorbs addition and multiplication: 
\begin{equation}\label{LIM3}
\aleph_0 = \aleph_0 + 1 = \aleph_0 - 1 = \aleph_0 \times \aleph_0
\end{equation}
and that $\aleph_0 - \aleph_0$  and $\frac{\aleph_0}{\aleph_0}$ are undefined. From the  definition \ref{SMALLINF} and equations \ref{LIMAL1}, \ref{LIMAL2}, \ref{LIMAL3}, \ref{LIMAL4} and \ref{LIMAL5} we can conclude that $\aleph_{-1}$ is more well-behaved: it still absorbs addition, but it is definite with respect to  multiplication and division. The real number $\aleph_{a-1}$ is infinite but smaller than  $\aleph_0$ since $\lim_{x \rightarrow \aleph_0} x^{\log x} = x$. This analysis shows that there are at least three types of infinity for number systems: 

 \begin{itemize}
 \item $\aleph_0$, the cardinality of $\mathbb{N}$, associated with \emph{counting}. 
 \item $\aleph_{a,1}$ the cardinality of $\mathbb{R}$, associated with the \emph{continuum}: $a^{\aleph_0}= \aleph_{a,1}$. 
 \item $\aleph_{a,-1} \in \mathbb{R}$ the amount of \emph{information} created when counting to infinity, measured by a logarithm with base $a$.~\footnote{We will omit the subscript $a$ when no confusion is possible.} We have: $\log_a (\aleph_{0}) = \aleph_{a,-1}$
 \end{itemize} 
 
 Now that we have an elementary calculus for $\aleph_{-1}$ we can give an interpretation: 

\begin{definition}[Information Limit]\label{INFLIM}
\[\lim_{n \rightarrow \aleph_0} I(n) = \lim_{n \rightarrow \aleph_0} \log n = \aleph_{-1}\]
\end{definition}

All counting processes produce, within a constant factor, the same amount of information in the limit. This gives rise to a theory of transfinite information measurement.~\footnote{ This definition opens up a whole potential universe of small infinities that mirrors the Cantor hierarchy, e.g. $\aleph_{-2}$ measures the size of numbers measuring the size of $\aleph_0$ with the conversion laws: $\log (\aleph_{n}) = \aleph_{n -1}$ and $2^{\aleph_{n}}= \aleph_{n+1}$. This falls beyond the scope of this paper.} Note that $\aleph_{-1}$ is conceptually different from $\aleph_0$ just as it is from $\aleph_1$. It is a \emph{scalar} value in $\mathbb{R}$ that represents a distance measurement and not a cardinality. The full power of definition \ref{SMALLINF} is clear from the following information limits: 
\begin{lemma}[Elementary Information Limits]\label{ELEMINFLIM}

 \begin{equation}\label{ELIM1}
   \lim_{x, y \rightarrow \aleph_0} I(x) + I(y) =  \lim_{x \rightarrow \aleph_0} I(x) + I(x) = \aleph_{-1} = O(\aleph_{-1})
   \end{equation}
   \begin{equation}\label{ELIM3}
\lim_{x, y \rightarrow \aleph_0} \frac{I(x)}{I(y)} = \lim_{x \rightarrow \aleph_0} \frac{I(x)}{I(x)}= O(1)
\end{equation}
 \begin{equation}\label{ELIM4}
\lim_{x, y \rightarrow \aleph_0} I(x) \times I(y) = \lim_{x \rightarrow \aleph_0} I(x) \times I(x) =O((\aleph_{-1})^2)
\end{equation}
\end{lemma}
Proof: Immediate consequence of definition \ref{SMALLINF} and equations \ref{LIMAL1}, \ref{LIMAL2}, \ref{LIMAL3}, \ref{LIMAL4}. We give one example:    
\[\lim_{x, y \rightarrow \aleph_0} I(x) + I(y) = \lim_{x, y \rightarrow \aleph_0} \log (x + 1 ) + \log(y + 1)
   = \aleph_{-1} + \aleph_{-1} = O( \aleph_{-1}) \]
   $\Box$

\section{Appendix: Sets of Incompressible numbers}\label{SETSINCOMP}

It is not trivial to estimate the exact information content of finite sets of numbers in general. This issue will be addressed in paragraph \ref{INFINNUMBERS}.  There is by definition no method to select or identify random numbers but the following lemmas provide a useful set of tools to work with random numbers:

\begin{definition}[Bounded Random Set]\label{BOUNDRANDSET} 
A \emph{bounded random set} $\{x_1, x_2,\dots, x_k\}$ with bound $n$ consists of a random selection of $k$ numbers from the set $\{0,1,2,\dots,n-1\}$ such that each number has a probability of $1/n$ to be selected. 
\end{definition}

\begin{lemma}[Incompressibility Finite Random Sets]\label{RANDOMSETS}
For any finite $k$, the density of incompressible sets of numbers of cardinality $k$ in the total set of sets of numbers with cardinality $k$ is $1$ in the limit. 
\end{lemma}
Proof: Immediate consequence of theorem \ref{INCOMPRESSIBILITY} and definition \ref{BOUNDRANDSET}. The density of the compressible numbers goes to zero in the limit. A bounded random set $\{x_1, x_2,\dots, x_k\}$ that consists of a random selection of $k$ numbers from the set $\{0,1,2,\dots,n-1\}$ such that each number has a probability of $1/n$ to be selected is incompressible with probability $1$ if $n$ goes to infinity. $\Box$

Random sets are a useful tool in proofs, because they allow us to formulate statements about the relative amount of information in expressions, even if we do not have a good theory of measurement. We can never select one deterministically, but we know that they must exist in abundance. 

\begin{lemma}[Relative Information for Random Sets]\label{RELINFSETS}
If  $A= \{x_1,x_2,\dots,x_k\}$ is a random set of natural numbers then 1) for every $B \subset A$ we have $I(B) < I(A)$ and 2)  If the elements are ordered as a k-tuple such that $x_1 < x_2 < ,\dots, < x_k$ then $I((x_1,x_2,\dots,x_k)) = \Sigma_{i=1}^k \log x_i$. 
\end{lemma}
Proof: 1) is an immediate consequence of the incompressibility of the elements of $A$. 2) Immediate consequence of definition \ref{FIRSTLAW} and the incompressibility of $\{x_1,x_2,\dots,x_k\}$. $\Box$

\begin{lemma}[Expected Values for Bounded Random Sets]\label{EXPVALBOUNDRANDSET}
The expected value of a random variable bounded by $n$, according to definition \ref{BOUNDRANDSET}, in the limit is $\mu_x= x_1p_1 + x_2p_2 + \dots + x_np_n= \Sigma x_ip_i =  \frac{1/2n(n+1)}{n} = 1/2(n+1)$.  If the set is sparse we may assume that the elements are drawn independently. The expected sum of a set with $k$ elements then is $1/2k(n+1)$. The expected product is $1/2(n+1)^k$.
\end{lemma}

A useful related concept is: 
\begin{definition}[Information Neighborhood Set]
An information neighborhood set for  $n \in \mathbb{N}$ and $\epsilon \in \mathbb{R}$ is $N_n^{\epsilon} = \{x \in \mathbb{N}|\log n < \log x < \log n + \epsilon\}$ 
\end{definition}
We can prove: 

\begin{lemma}[Maximal Entropy Neighborhood Lemma]\label{ENTROPYLEMMA}
For any $\epsilon$ as $n$ goes to infinity the density of k-subsets of the information neighborhood set $N_n^{\epsilon}$ that only contain incompressible numbers goes to $1$. 
\end{lemma}
Proof: immediate consequence of theorem \ref{INCOMPRESSIBILITY}. In the limit the size of the set $\{x \in \mathbb{N}|\log n < \log x < \log n + \epsilon\}$ goes to infinity, consequently the density of compressible numbers goes to zero. So when we select $k$ numbers the possibility that one of them is compressible in the limit is zero. $\Box$

The lemma tells us that in the limit we can find any amount of incompressible sets of incompressible numbers in any information neighborhood of a given number. Lemma \ref{EXPVALBOUNDRANDSET} tells us that random bounded sets tend to `cluster together' in the limit and thus exist in the same information neighbourhood.

\section{Appendix: Counting Sets by Cardinality ordered by their Sums}\label{COUNTSETSUM}

\begin{figure}[!t]
\centering
\fbox{\includegraphics[ width=320bp]{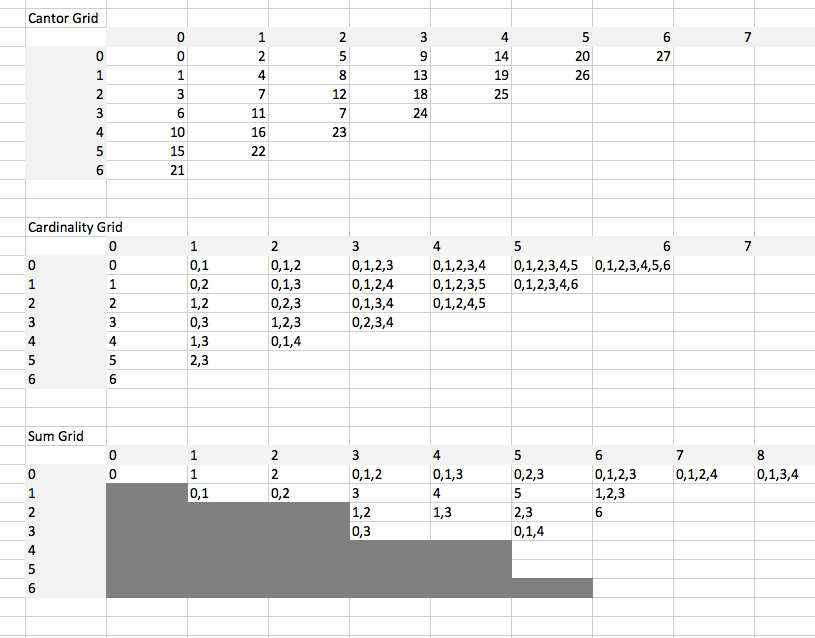}}
\caption{In this figure we see three fragments of grid functions: the Cantor grid $\mathcal{G}^2(\mathbb{N})$, the Cardinality Grid $\mathcal{G}^2_{card}( \mathfrak{P}(\mathbb{N}))$  where the elements of $\mathfrak{P}(\mathbb{N})$ are stored lexicographically in bins of the same cardinality, and the corresponding Sum Grid, $\mathcal{G}^2_{sum}( \mathfrak{P}(\mathbb{N}))$ where the elements of $\mathfrak{P}(\mathbb{N})$ are distributed over bins of the same sum using the ordering defined in $\mathcal{G}^2_{card}( \mathfrak{P}(\mathbb{N}))$.The elastic shift is clearly visible, the shaded cells are vacuous.}
\label{GRIDEXAMPLES}
\end{figure}

We have established the rigidity of $\mathcal{G}^2_{card}( \mathfrak{P}(\mathbb{N}))$, here all sets with the same cardinality end up in the same columns or bin. We now investigate whether we could use another binning function based on addition: $\forall (x \in  \mathfrak{P}(\mathbb{N})) bin_{sum}(x)=\Sigma_{x_i \in x} x$.  Since it is not clear that there exists a straightforward function that defines the set, we specify the following construction: 

\begin{definition}[Cantor grid on sum] \label{CANTORGRIDSUM}
The object $\mathcal{G}^2_{sum}( \mathfrak{P}(\mathbb{N}))$ is defined by the  infinite \emph{distribution algorithm} $h_{\texttt{sum}}:  \mathfrak{P}( \mathbb{N}) \rightarrow  \mathbb{N} \times \mathbb{N}$: 
\begin{itemize}
\item For each $i \in \mathbb{N}$ there is a bin $S_i$. Initially all the bins are empty. 
\item We enumerate $\mathfrak{P}(\mathbb{N})$ using $\mathcal{G}^2_{card}( \mathfrak{P}(\mathbb{N}))$ and assign each set $s$ to a location $(k, j)$, where $k = \Sigma_{x_i \in x} x_i$ is a bin, and $j$ the first free index in bin $k$. 
\end{itemize} 
\end{definition}

The procedure is illustrated in figure \ref{GRIDEXAMPLES}. We make some observations about $\mathcal{G}^2_{sum}( \mathfrak{P}(\mathbb{N}))$: 
\begin{itemize}
\item The procedure uses an infinite number of variables that are updated incrementally, i.e. during its construction an unbounded amount of information is created.    
\item The number of sets in $\mathfrak{P}(\mathbb{N})$ that add up to $k$ is finite, so the bins in $\mathcal{G}^2_{sum}( \mathfrak{P}(\mathbb{N}))$ will only have indexes on a finite initial segment. $h_{\texttt{sum}}$ is an injection. The rest of the indexes in the bin are not used in the mapping. We'll call these indexes \emph{vacuous} and the others \emph{non-vacuous}.
\item During the construction process the points are distributed over the bins like a cloud. At each stage of the process the exact contents of an initial segment of the function is known. This allows us to make density estimates for the non-vacuous cells. 
\item Because of the existence of vacuous indexes there can never be a one to one correspondence between $\mathcal{G}^2_{card}( \mathfrak{P}(\mathbb{N}))$ and $\mathcal{G}^2_{sum}( \mathfrak{P}(\mathbb{N}))$, since  the index sets in  $\mathcal{G}^2_{card}( \mathfrak{P}(\mathbb{N}))$ are infinite. This enumeration will enlist both vacuous and non-vacuous indexes. 
\end{itemize}

\begin{lemma}\label{EXPTIME}
The function $h_{\texttt{sum}}$ can be computed in exponential time and polynomial space. 
\end{lemma}
Proof: Just emulate the construction, counting, but forgetting, the bins that are not needed. If $h_{\texttt{sum}}(s)=(k,j)$ then $s$ is located in bin $k$ at index $j$. It can be computed in exponential time and polynomial space. Such an algorithm has a \emph{set counter} $i$ that enumerates the elements of $\mathfrak{P}(\mathbb{N})$ and a \emph{bin counter} counting the number of times an element is added to bin $k$ before $j$ is selected. $\Box$

\begin{figure}[!t]
\centering
\fbox{\includegraphics[ width=250bp]{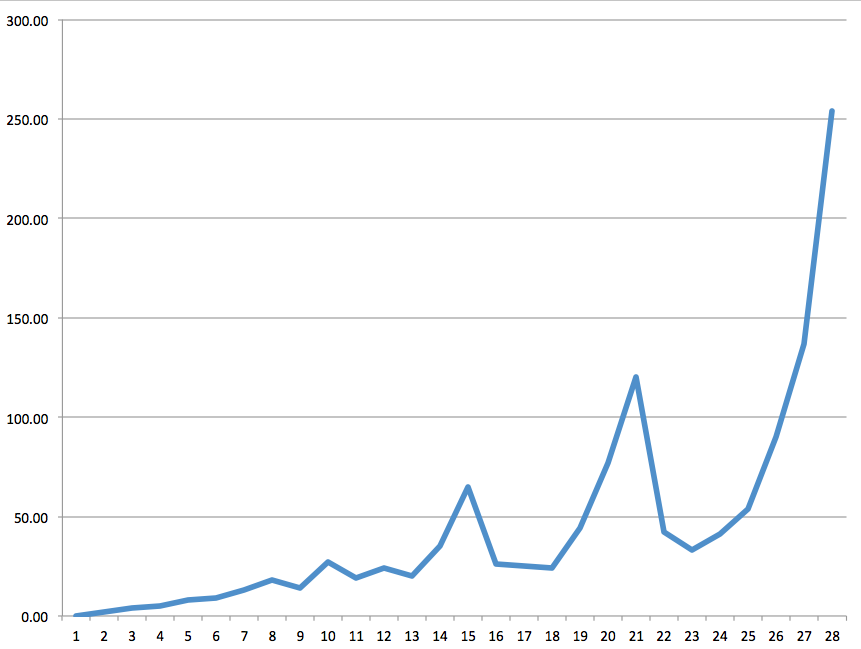}}
\caption{The \emph{information graph for addition of finite sets of numbers} for objects up to 5 bits.. It encodes the information about the function "the $n^{th}$ set of numbers that adds up to $k$" for the first 28 numbers in $\mathbb{N}$. Inspection of this small initial segment of the graph already shows its erratic behavior. The information in this graph grows incrementally with its construction and is infinite in the limit.}
\label{SUMGRIDPLOT}
\end{figure}

 The information graph for addition   \ref{SUMGRIDPLOT}  encodes the information about the function "the $n^{th}$ set of numbers that adds up to $k$". Note that all transformations involved, except $h_{\texttt{sum}}$, have polynomial time complexity. This graph has some peculiar qualities, with interesting mathematical consequences. Visual inspection already suggests that the graph does not define a rigid mapping. Note that we need to select values for $\mathcal{G}^2_{sum}( \mathfrak{P}(\mathbb{N}))$ that stay as close as possible to the enumeration defined bij $\mathcal{G}^2_{card}( \mathfrak{P}(\mathbb{N}))$, which is our measuring tool. That is why we used the ordering of $\mathcal{G}^2_{card}( \mathfrak{P}(\mathbb{N}))$ to enumerate the set. Each set $s$ of cardinality $k$ with index $(k,j)$ in $\mathcal{G}^2_{card}( \mathfrak{P}(\mathbb{N}))$ is shifted to the right to at least bin $1/2 k(k-1)$  in $\mathcal{G}^2_{sum}( \mathfrak{P}(\mathbb{N}))$. It will get a new location $(p,q)$ where $p \geq 1/2 k(k-1)$ and $q < j$. It would be surprising if this new index encodes in $\mathcal{G}^2_{sum}( \mathfrak{P}(\mathbb{N}))$ exactly the information contained in $s$, as in the case in $\mathcal{G}^2_{card}( \mathfrak{P}(\mathbb{N}))$.

The density per bin of the non-vacuous (white) cells is zero in the limit when counted per bin, but approximates one when enumerating $\mathcal{G}^2_{sum}( \mathfrak{P}(\mathbb{N}))$ diagonally via the Cantor function. The number of white cells grows exponentially with the amount of sets of numbers that add up to $k$.  The main result of this paper implies that there is no polynomial function that gives us direct access to the points in the last plot of figure \ref{GRIDEXAMPLES}.

\section{Appendix: All polynomials are unbounded}\label{POLUNBOUND}

\begin{theorem}\label{POLUNBOUNDLEM}
All polynomial functions are unbounded in the limit. 
\end{theorem}
Proof: 
\begin{itemize}
\item Case 1) The theorem holds for all univariate polynomials $p(x)=a_{n}x^n+a_{n - 1}x^{n - 1}+ \dots +a_1x+a_0$. When $x \rightarrow \infty$  we have $p(x) \sim a_nx^n$ which goes to infinity. 
\item Case 2) The theorem holds for all linear multivariate polynomials $p=a_nx_n+a_{n - 1}x_{n - 1}+\dots+a_1x+a_0$. For each $c$ the set of solutions to the diophantine equation $p=c$ defines an $n-1$ dimensional hyperspace, so $p$ is not bounded by a $c$. 
\item Case 3) Suppose a polynomial $p$, of $k$ variables of degree $m$. Let $Q$ be the multiset of terms in $P$ with the highest degree $m$. By the same argument as in case 1) only those terms are important in the limit. Form the corresponding polynomial $q$: 

\begin{equation}\label{EQA1}
q = c_1x_1^{a_1}x_2^{b_1}x_3^{c_1} . . . x_k^{k_1}
+ 
c_2x_1^{a_2}x_2^{b_2}x_3^{c_2} . . . x_k^{k_2}
+ \dots + 
c_ix_1^{a_i}x_2^{b_i}x_3^{c_i} . . . x_k^{k_i}
\end{equation} 
  
For all the terms in $q$ we have  $(a_i + b_i + c_i, . . . + k_i) = m$. If $m=1$ then $p$ is linear (case 2). 

Suppose $m>1$.  Rewrite q as the set of equations: 
\[y_1= x_1^{a_1}x_2^{b_1}x_3^{c_1} . . . x_k^{k_1}\]
\[y_2 =x_1^{a_2}x_2^{b_2}x_3^{c_2} . . . x_k^{k_2}\]\[  \dots \] 
\[y_i=c_ix_1^{a_i}x_2^{b_i}x_3^{c_i} . . . x_k^{k_i}\]
\begin{equation}\label{EQA2}
q' = c_1y_1 +  c_2y_2+ \dots + c_iy_i
\end{equation} 

By lemma \ref{EXPVALBOUNDRANDSET} the expected max-entropy value of the term $c_jx_1^{a_j}x_2^{b_j}x_3^{c_j} . . . xk^{k_j} = c_jy_j= c_i(1/2(n+1))^k$. By lemma \ref{ENTROPYLEMMA} in the limit the  density of independent incompressible number in any neighborhood of $x_1^{a_i}x_2^{b_i}x_3^{c_i} . . . x_k^{k_i}= 1/2(n+1))^k$ goes to one. So we may interpret the density of solutions for equation \ref{EQA1} in the limit to be equal of that of a linear equation \ref{EQA2}, which is case 1). 
\end{itemize}
$\Box$

\end{document}